# Numerical analysis of the spectrum of the Dirac operator in four-dimensional SU(2) gauge fields *

Thomas Kalkreuter[a] **

[a]Institut für Physik, Humboldt-Universität, Invalidenstraße 110, D–10099 Berlin, Germany
Email: kalkreut@linde.physik.hu-berlin.de

Two numerical algorithms for the computation of eigenvalues of Dirac operators in lattice gauge theories are described: one is an accelerated conjugate gradient method, the other one a standard Lanczos method. Results obtained by Cullum's and Willoughby's variant of the Lanczos method (whose convergence behaviour is closely linked with the local spectral density) are presented for euclidean Wilson fermions in quenched and unquenched SU(2) gauge fields. Complete spectra are determined on lattices up to $8^3 \cdot 12$, and we derive numerical values for fermionic determinants and results for spectral densities.

## 1. INTRODUCTION

In order to study questions of chiral symmetry breaking [1–6] and universality [7,8], and also in the context of Lüscher's fermion algorithm [9–11], one is interested in the eigenvalues of ($\gamma_5$ times) the gauge covariant Dirac operator which are close to the origin. In this talk we focus on the numerical aspects of the determination of spectral properties of lattice Dirac operators. Two versions of lattice fermions are in use today [12]: Wilson fermions [13] and staggered fermions [14,15].

Low-lying eigenvalues of staggered fermions were investigated e.g. in [2]. The present author obtained complete spectra [16], and these data were further analysed by Halasz and Verbaarschot [8]. They were interested in universal fluctuations in spectra of lattice Dirac operators, and they verified that the correlations among the eigenvalues of staggered fermions in SU(2) gauge fields are described by the Gaussian symplectic ensemble of random matrix theory with the chiral symmetry of the Dirac operator built in.

Eigenvalues of Wilson fermions were computed e.g. in Refs. [17,3], and recently in [18]. Some of the latter results will be presented in this talk where we consider the hermitean operator

$$Q = \gamma_5 \, (D+m) \, / \, (8+m) \tag{1}$$

with periodic boundary conditions. $Q$ is normalized such that its eigenvalues are between -1 and 1. $(D+m)$ is the Dirac operator for Wilson fermions of bare mass $m$. On a four-dimensional lattice $\Lambda$ of sites $x$ (with lattice spacing $a=1$), $(D+m)$ acts on a lattice spinor $\psi$ as follows,[1] see e.g. [12],

$$[(D+m)\psi]^{\alpha a}(x) = \frac{1}{2\kappa}\, \psi^{\alpha a}(x)$$
$$-\frac{1}{2}\sum_{\mu=1}^{4} \{ (\mathbb{1}-\gamma_\mu)_{\alpha\beta}\, U(x,x+\mu)_{ab}\, \psi^{\beta b}(x+\mu)$$
$$+ (\mathbb{1}+\gamma_\mu)_{\alpha\beta}\, U(x,x-\mu)_{ab}\, \psi^{\beta b}(x-\mu) \}. \tag{2}$$

Here $\kappa = (2m+8)^{-1}$ denotes the hopping parameter and $x \pm \mu$ is the nearest neighbour site of $x$ in $\pm\mu$-direction. The gauge field $U(x, x \pm \mu) \in \mathrm{SU(N_c)}$ lives on the links $(x, x\pm\mu)$ of the lattice and is generated by some Monte Carlo process [12]. On the rhs of eq. (2) an implicit summation over the spinor indices ($\beta = 1, \ldots, 4$) and colour ($b = 1, \ldots, N_c$) is understood.

Two algorithms for eigenvalues will be described in Sec. 2, and results in four-dimensional SU(2) gauge fields will be presented for complete spectra of Wilson fermions in Sec. 3.

*to appear in Nucl. Phys. B (Proc. Suppl.), Proceedings of the International Symposium Ahrenshoop on the Theory of Elementary Particles, Buckow'95.
**Work supported by Deutsche Forschungsgemeinschaft, grant Wo 389/3-1.

[1]The hermitean euclidean $\gamma$-matrices $\gamma_\mu$, $\mu = 1,2,3,4$, satisfy the Clifford algebra $\{\gamma_\mu, \gamma_\nu\} = 2\delta_{\mu\nu}\mathbb{1}$. Moreover, $\gamma_5 \equiv \gamma_1\gamma_2\gamma_3\gamma_4$ anticommutes with all of them and $\gamma_5^2 = \mathbb{1}$ is the $4 \times 4$ unit matrix.

## 2. ALGORITHMS FOR EIGENVALUES

Maybe the most obvious candidate method which comes to mind to numerically determine eigenvalues of very large (hermitean or anti-hermitean) matrices[2] is a Lanczos algorithm [19] which we explain in Sec. 2.2. Variants of this method have been used in lattice field theory for a long time, see e.g. Refs. [20,2] for staggered fermions and Refs. [17,3] for Wilson fermions.

In order to obtain an overview of complete spectra the Lanczos method is well-suited in lattice QCD, as we will see in Sec. 3. However, the method can be problematic whenever one is interested in only a few eigenvalues. Whether eigenvalues have converged can be estimated only from experience [18], and not from a rigorous error bound.[3] Moreover, a Lanczos method cannot provide information about multiplicities.

Some applications also require knowledge of the eigenvectors, for instance to isolate the contribution of low-lying eigenmodes to physical observables. In such cases a conjugate gradient (CG) algorithm is superior, and even if one just wants knowledge of eigenvalues the CG method is favourable because it does not suffer from the drawbacks and potential pitfalls of the Lanczos method. So let us turn first to an accelerated version of a straightforward CG implementation [22].

### 2.1. Accelerated CG method

In [23] a CG method was proposed for the computation of low-lying eigenvalues of $Q^2$, cf. also [24,25]. More generally, this procedure can be used to obtain extreme eigenvalues of a hermitean operator. The CG method is based on the gauge covariant extremization of the Ritz functional[4]

$$\mu(\psi) = \frac{\langle \psi, Q^2 \psi \rangle}{\langle \psi, \psi \rangle} \qquad (\psi \neq 0) \ , \tag{3}$$

where for the $k$-th lowest eigenvalue $\psi$ is kept orthogonal to the eigenspace of the $(k-1)$ lower eigenvalues. The pure CG minimization of Ref. [23] was accelerated in [22] by alternating CG searches with exact diagonalizations in the subspace spanned by the numerically computed eigenvectors. This method is attractive because of the following key features.

- Rigorous error bounds for eigenvalues can be derived just from the last CG iterate.
- The correct multiplicities are detected.
- Approximations to eigenvectors are obtained as a by-product.
- The pure CG algorithm is speeded up through the intermediate diagonalizations by a factor of $4-8$.

Furthermore, the algorithm is numerically very stable, even if one is restricted to single precision arithmetics, and it can be parallelized in an efficient way. For more details we refer to Ref. [22].

### 2.2. Lanczos method

The Lanczos procedure is a technique that can be used to solve large, sparse, symmetric or hermitean eigenproblems[5] [19]. The idea is to transform a given hermitean $n \times n$ matrix $A$ into a similar symmetric tridiagonal matrix $T = V^{-1}AV$ with unitary $V$, and then $T$ is diagonalized. The transformation of $A$ can be performed iteratively. If one writes $V = (v_1, v_2, \ldots, v_n)$ with column vectors $v_i$ ("Lanczos vectors") and

$$T = \begin{pmatrix} \alpha_1 & \beta_1 & & \\ \beta_1 & \alpha_2 & \ddots & \\ & \ddots & \ddots & \beta_{n-1} \\ & & \beta_{n-1} & \alpha_n \end{pmatrix} \ , \tag{4}$$

then $AV = VT$ is equivalent to $(i = 2, \ldots, n-1)$,

$$\begin{aligned} Av_1 &= \alpha_1 v_1 + \beta_1 v_2 \ , \\ Av_i &= \beta_{i-1} v_{i-1} + \alpha_i v_i + \beta_i v_{i+1} \ , \\ Av_n &= \beta_{n-1} v_{n-1} + \alpha_n v_n \ . \end{aligned} \tag{5}$$

Given an initial — generally random — vector $v_1$ and using the orthonormality among the $v_i$ with respect to the canonical scalar product $\langle \cdot \, , \cdot \rangle$, one can determine iteratively from these equations:

[2] $Q$ is represented by an $n \times n$ matrix with $n = 4N_c|\Lambda|$.
[3] In principle one can quote error bounds [19,21], but they are not practical in lattice gauge theories.
[4] $\langle \cdot \, , \cdot \rangle$ denotes the scalar product of the Hilbert space.
[5] In Refs. [20,17,3] a non-hermitean Lanczos method was used and it was concluded that it works on small lattices.

$\alpha_i = \langle v_i, Av_i \rangle$, $\beta_i^2 = (\langle v_i, A^2 v_i \rangle - \alpha_i^2 - \beta_{i-1}^2)$, (with $\beta_0 \equiv 0$), and $v_{i+1} = \beta_i^{-1}(Av_i - \beta_{i-1}v_{i-1} - \alpha_i v_i)$ as long as $\beta_i \neq 0$ (otherwise the iteration is stopped). $T$ will be a real matrix also in case that $A$ is complex, and the sign of $\beta_i$ is arbitrary.

In exact arithmetic, the Lanczos iteration should finish after at most $n$ steps and the last equation in (5) would be automatically fulfilled. For this case there exists a convergence theory [19]. In practice, however, there are severe problems with a straightforward implementation [19,21]. These problems are caused by rounding errors and loss of orthogonality among the Lanczos vectors.[6] In principle the latter problem can be circumvented by storing all the Lanczos vectors and enforcing orthogonality among them by hand. But then one is restricted to small lattices because of computer memory or I/O limitations.

A practical Lanczos method is due to Cullum and Willoughby [21]. In their proposal one performs no reorthogonalization, and one continues the iteration (5) for an a priori unspecified count. In this way a sequence of $j \times j$ tridiagonal matrices $T^{(j)}$, $j = 1, 2, \ldots$ is generated. There exists a recipe [21] how the eigenvalues of $T^{(j)}$ are connected with eigenvalues of $Q$, see [18] for details.

## 3. NUMERICAL RESULTS

Cullum's and Willoughby's Lanczos method [21] was studied in quenched and unquenched[7] SU(2) gauge fields, i.e. $N_c = 2$ in the following. Tridiagonal matrices were diagonalized by means of the Numerical Recipes routine "tqli" [26] which implements the QL algorithm with implicit shifts. Two eigenvalues $\lambda$ were counted as different when they differed by more than $10^{-10}$. This number is arbitrary but it is chosen such that it is small compared with the gaps in the spectra, and large compared with round-off errors. The computer program was checked for gauge covariance, and it was also verified that free spectra are obtained correctly, except for multiplicities.

In the following we present results for complete spectra. Further results and a comparison of the partially converged Lanczos method with the accelerated CG algorithm for low-lying eigenvalues of Ref. [22] can be found in [18].

[6] This loss of orthogonality is not necessarily due to the accumulation of round-off errors [19].

[7] With two flavours of dynamical fermions.

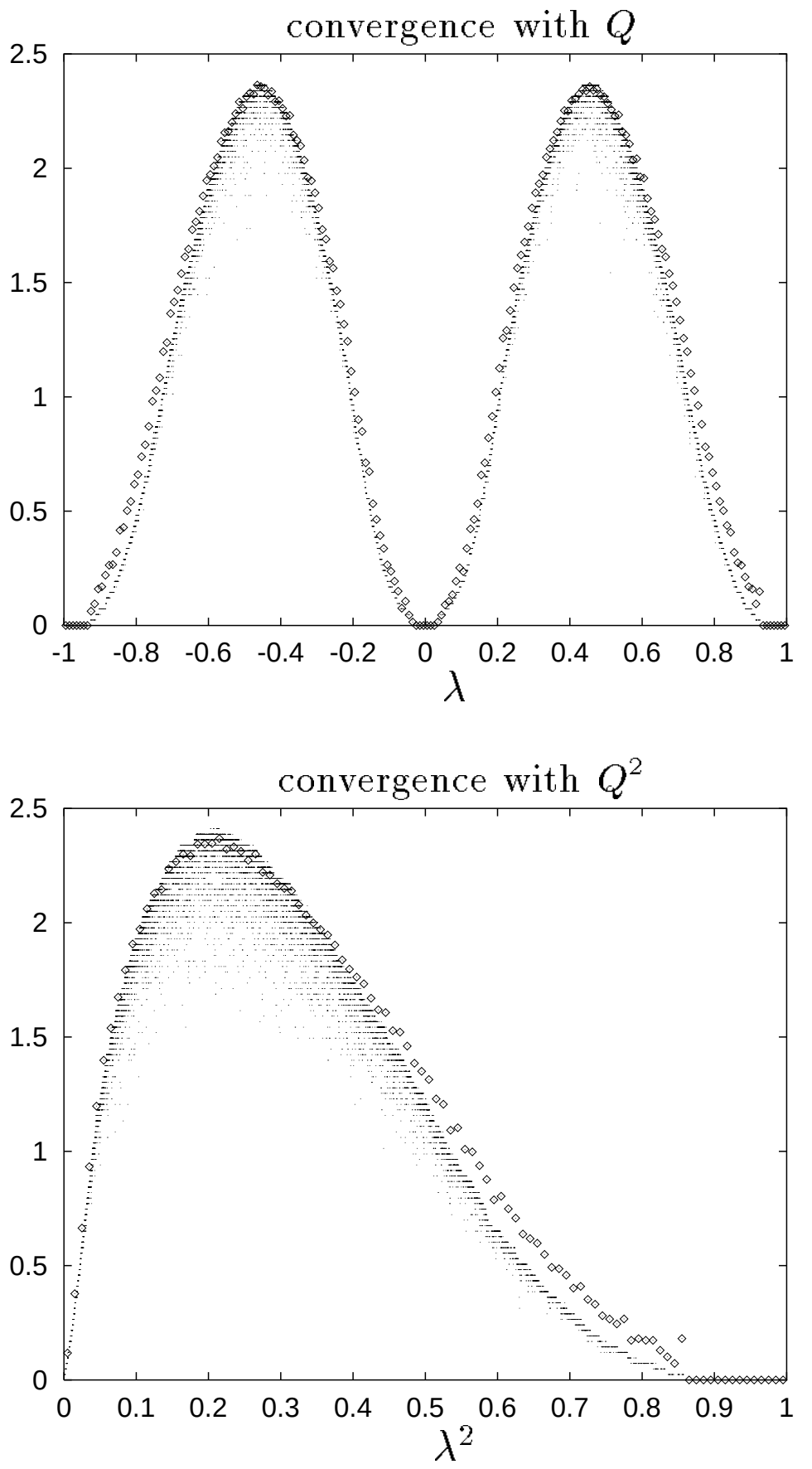

Figure 1. Convergence of the Lanczos method.

### 3.1. Convergence behaviour

Let us start by looking at the convergence behaviour of the Lanczos algorithm. We would like to concretize what is meant by the general statement of Ref. [21] that the local gap structure plays a role. An example with $n = 20\,736$ eigenvalues in an unquenched configuration on a $6^3 \cdot 12$ lattice at[8] $\beta = 2.12$, $\kappa = 0.15$ is shown in Fig. 1. The Lanczos method was run both with $A = Q$ and with $A = Q^2$. Every 500 iterations convergence was checked by a comparison with reference data from $T^{(82\,944)}$. Dots give the number

[8] Here and in the following $\beta$ denotes the coupling constant of the gauge part of the action, see e.g. [12].

$j$ of Lanczos iterations in multiples of $n$ which are required for convergence of $\lambda$ or $\lambda^2$, and diamonds indicate $1.75[\lambda_{\max} - \lambda_{\min} - |\lambda|]\rho(\lambda)$ and $3.5[(\lambda_{\max} - \lambda_{\min})\lambda - \lambda^2]\rho(\lambda^2)$, where $\lambda_{\max/\min}$ denotes the highest/lowest eigenvalue of $Q$.

We notice that the number of Lanczos iterations required for convergence is intimately related with the local spectral density. For this reason it is advantageous to diagonalize the unsquared operator in case that one requires the lowest or highest eigenvalues of $Q^2$. This statement will become more and more significant as a critical point is approached in the $\beta$-$\kappa$-plane.

### 3.2. Complete spectra

Complete spectra were determined in the configurations quoted in Tables 1 and 2 below. In nontrivial gauge fields no degeneracies were found, neither for $Q$ nor for $Q^2$. In case of $Q$ we always encountered an equal number of negative and positive eigenvalues. Furthermore, no discrepancies were found when the squared eigenvalues of $Q$ were compared with the results for $Q^2$.

Examples for complete spectra in a quenched gauge field at $\beta = 1.8$ on a $6^4$ lattice are given in Fig. 2. The integrated densities of eigenvalues $N(\lambda)$ and $N(\lambda^2)$ follow directly from the numerical data. They are normalized such that they take values between zero and one. At $\kappa = 0.15$ curves of a quenched $12^4$ gauge field[9] at $\beta = 1.8$ and unquenched gauge fields at $\beta = 2.12$ on $6^3 \cdot 12$ and $8^3 \cdot 12$ lattices coincide (on the scale of the figure) with the result on the $6^4$ lattice.

We have the following consistency checks which provide good evidence that all computed complete spectra are correct. First, on all investigated $4^4$ – $8^3 \cdot 12$ lattices the correct number of $n$ different eigenvalues was found. Second, we have analytical sum rules for powers of the eigenvalues, cf. also Ref. [16]. The traces (over colour and spinor indices) of $Q$ and $Q^3$ equal zero, and the trace of $Q^2$ reads in any unitary gauge field

$$\mathrm{Tr}\,\mathrm{Q}^2 = 4\,\mathrm{N_c}\,|\Lambda|\,(4 + \frac{1}{4\kappa^2})\,/\,(4 + \frac{1}{2\kappa})^2 \ . \qquad (6)$$

We obtained $|\mathrm{Tr}\,\mathrm{Q}| \lesssim 10^{-8}$, $|\mathrm{Tr}\,\mathrm{Q}^3| \lesssim 10^{-7}$, and

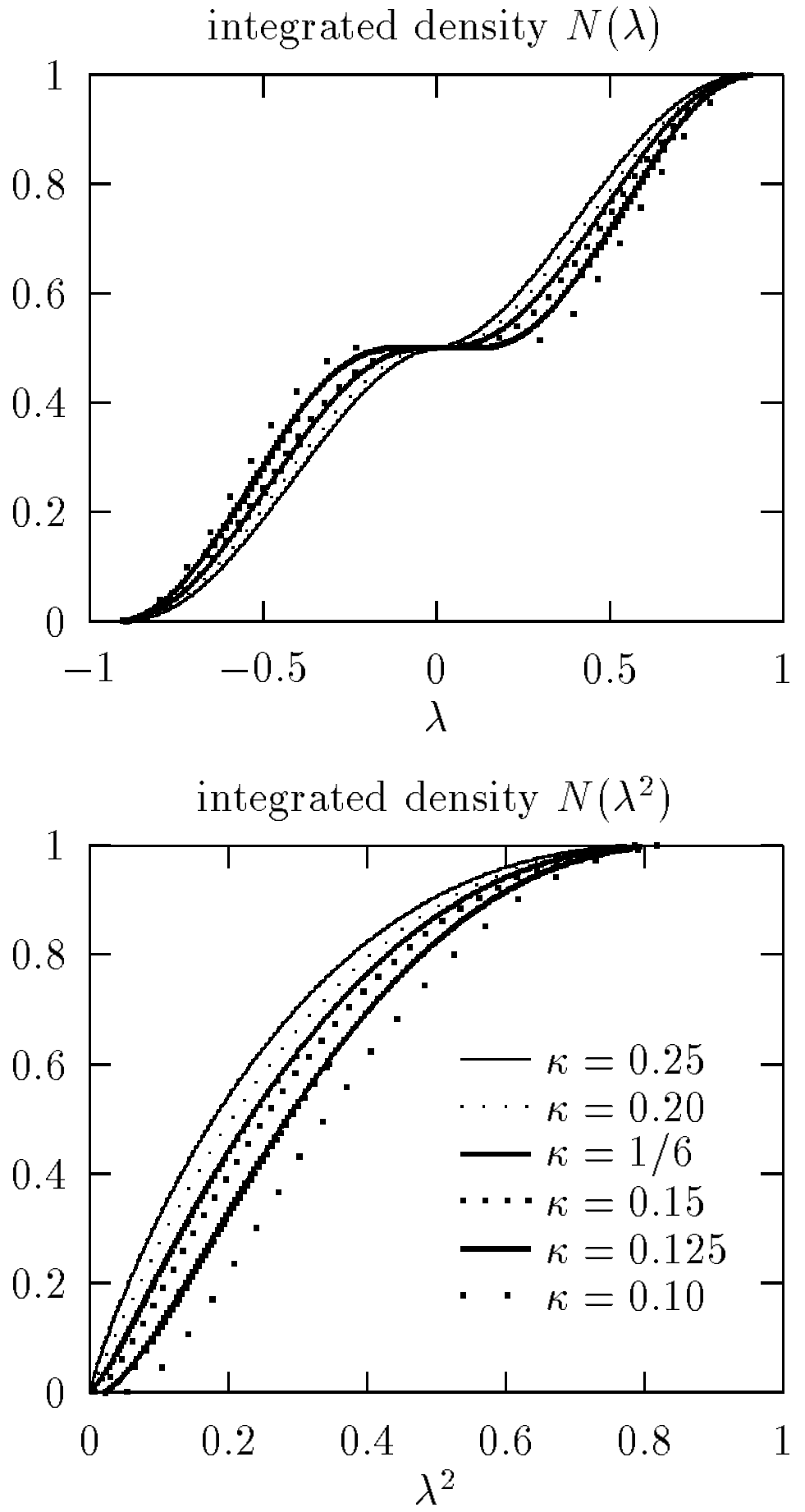

Figure 2. Integrated spectral densities.

$\mathrm{Tr}\,\mathrm{Q}^2$ came out with a relative accuracy of $10^{-8}$ – $10^{-12}$ (decreasing with increasing $|\Lambda|$). One could check for further sum rules by examining traces of higher powers of $Q$. Finally, we confirmed that the effort for the determination of complete spectra grows with $|\Lambda|^2$, $j = 4\,n$ worked in all cases.

### 3.3. Fermionic determinants

For free Wilson fermions spectra and determinants are known analytically [18]. Quenched / unquenched examples are contained in Table 1/2. One concludes a nice exponential dependence of the determinant on the lattice volume, already for relatively small lattices.

In a quenched Monte Carlo simulation the fermionic determinant is kept at a fixed value. From Table 2 one can get a feeling for the fluc-tuations of the determinant in an unquenched run. We can compare the values of eight independent dynamical $6^3 \cdot 12$ configurations at $\beta = 2.12$, $\kappa = 0.15$. The logarithmic entries in Table 2 fluctuate by $\approx 0.0003$ which translates to a fluctuation of the determinant itself by six orders of magnitude on the $6^3 \cdot 12$ lattice.

[9] Because of memory limitations, on the $12^4$ lattice all but four of the 165888 eigenvalues were found; see [18].

Table 1
Examples for Wilson fermions in quenched SU(2) gauge fields: $(\log_{10} \det Q)/(4N_c|\Lambda|)$.

| $\|\Lambda\|$ | $\beta$ | $\kappa = 0.1$ | $\kappa = 0.125$ | $\kappa = 0.15$ | $\kappa = 1/6$ | $\kappa = 0.20$ | $\kappa = 0.25$ |
|---|---|---|---|---|---|---|---|
| $6^4$ | 1.80 | -0.254757 | -0.299665 | -0.339289 | -0.362812 | -0.402336 | -0.440477 |
| $6^4$ | 2.80 | -0.254331 | -0.298278 | -0.334999 | -0.354646 | -0.383828 | -0.409753 |
| $6^4$ | 0.00 | -0.255279 | -0.301046 | -0.342455 | -0.368026 | -0.415065 | -0.477151 |
| $8^4$ | 0.00 | -0.255277 | -0.301041 | -0.342445 | -0.368010 | -0.415038 | -0.476955 |

Table 2
Examples for Wilson fermions in unquenched SU(2) gauge fields: $(\log_{10} \det Q)/(4N_c|\Lambda|)$.

| $\|\Lambda\|$ | $\beta$ | $\kappa = 0.15$ |
|---|---|---|
| $4^4$ | 1.75 | -0.338947 |
| $6^3 \cdot 12$ | 2.12 | -0.337169 |
| $6^3 \cdot 12$ | 2.12 | -0.337129 |
| $6^3 \cdot 12$ | 2.12 | -0.337282 |
| $6^3 \cdot 12$ | 2.12 | -0.337388 |
| $6^3 \cdot 12$ | 2.12 | -0.337385 |
| $6^3 \cdot 12$ | 2.12 | -0.337266 |
| $6^3 \cdot 12$ | 2.12 | -0.337269 |
| $6^3 \cdot 12$ | 2.12 | -0.337235 |
| $8^3 \cdot 12$ | 2.12 | -0.337295 |

### 3.4. Spectral densities

The density of eigenvalues of $Q^2$ around zero can be related with the chiral limit, if one connects the spectral density of $Q^2$ with that of the Dirac operator (not multiplied by $\gamma_5$) in the spirit of the "pion norm" of Ref. [1]. Such an analysis will however be done elsewhere. From the algorithmic point of view spectral densities $\rho(\lambda)$ of $Q$ and $\rho(\lambda^2)$ of $Q^2$ are interesting because they determine convergence properties (cf. Fig. 1).

Figs. 3 and 4 show normalized quenched $\rho$'s on a $6^4$ lattice at $\beta = 1.8$ and in a random $8^4$ gauge field. (Additional results at $\beta = 2.8$ are in [18].) The solid curves indicate reference data of unquenched gauge fields at $\kappa = 0.15$, $\beta = 2.12$ on $6^3 \cdot 12$ and $8^3 \cdot 12$ lattices, where results coincide.

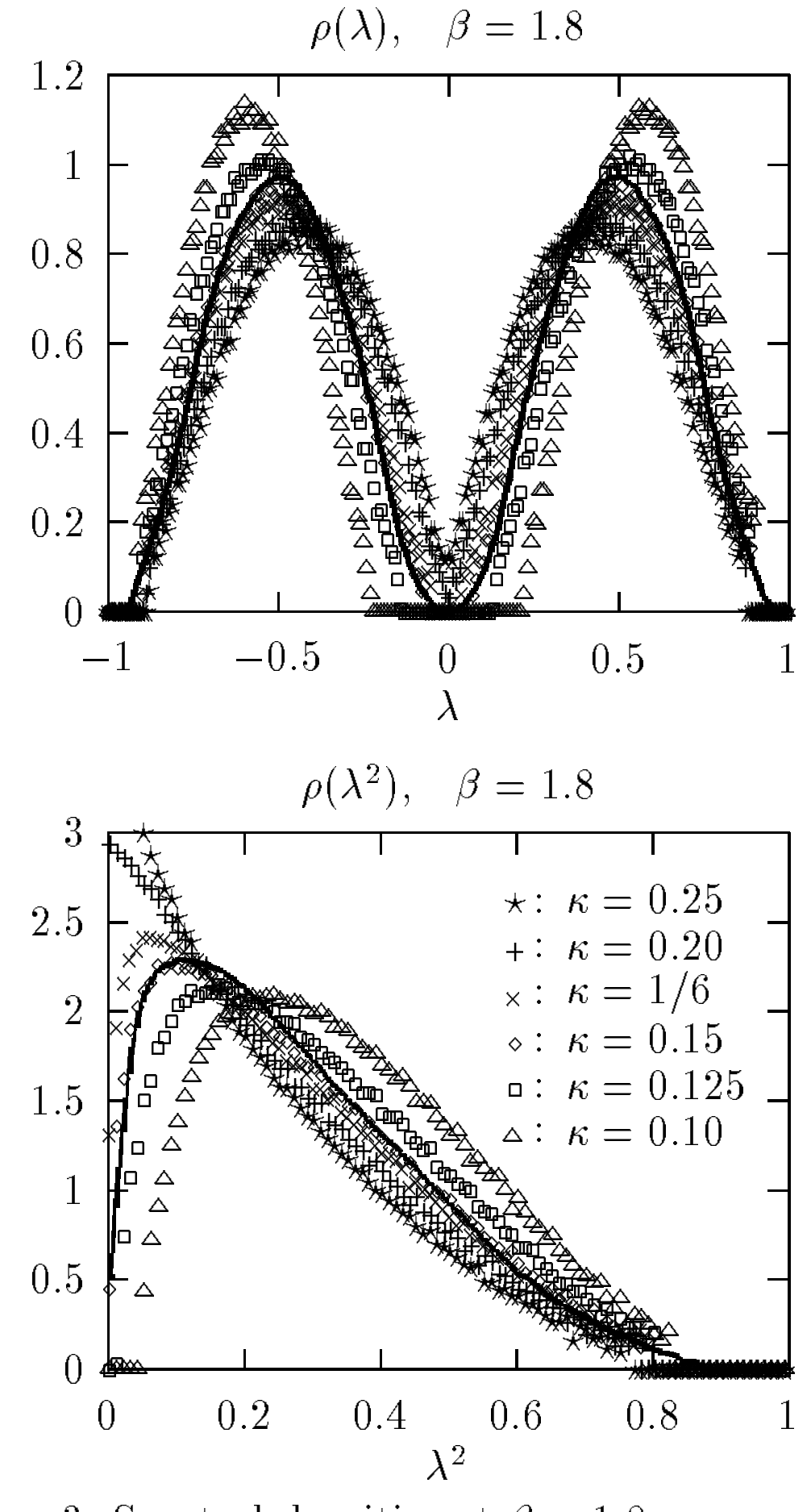

Figure 3. Spectral densities at $\beta = 1.8$.

The information in Figs. 3 and 4 is redundant because of the relation $\rho(\lambda^2) = \rho(\lambda)/(2\lambda)$, which can be verified numerically to a very high precision. Despite this redundancy we consider it worthwhile to present results for $\rho(\lambda)$ as well as for $\rho(\lambda^2)$. Note for instance that $\rho(\lambda^2)$ stays finite (must diverge) as $\lambda^2 \to 0$ when $\rho(\lambda = 0) = 0$ ($\rho(\lambda = 0)$ is finite).

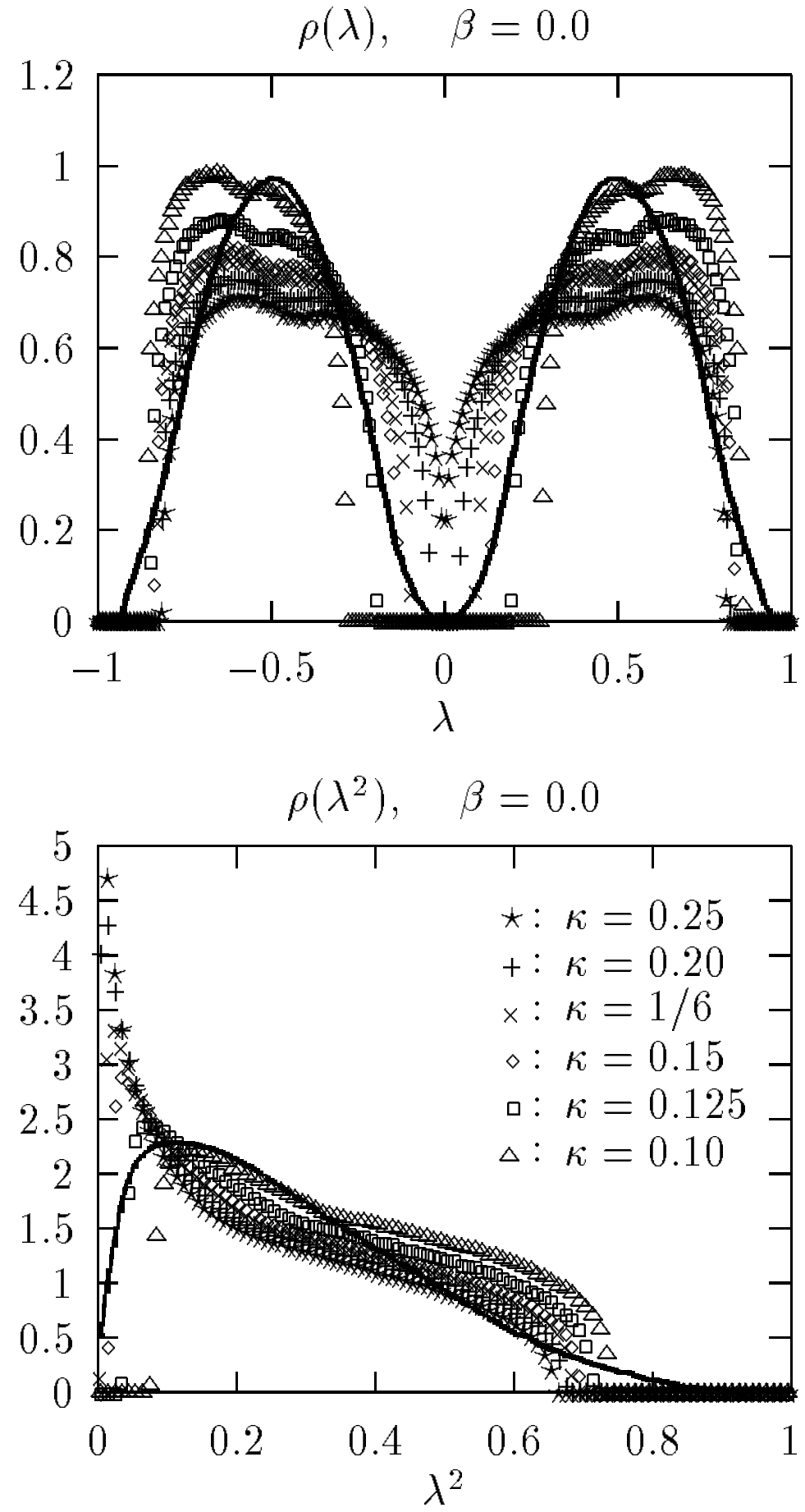

Figure 4. Spectral densities at $\beta = 0$.

I wish to thank B. Bunk, K. Jansen, H. Simma, and U. Wolff for discussions. C. Liu and K. Jansen are thanked for providing the unquenched Wilson configurations. Financial support by Deutsche Forschungsgemeinschaft under grant Wo 389/3-1 is gratefully acknowledged. The computations reported here were performed on HP workstations of the Humboldt-University.